# The silver-plated mirrors of the Historical Collection of Physics Instruments of Palermo University

Gli specchi argentati della Collezione Storica degli Strumenti di Fisica dell'Università di Palermo

**Giovanni Luca Sferrazza, Aurelio Agliolo Gallitto**

Dipartimento di Fisica e Chimica - Emilio Segrè, Università degli Studi di Palermo
via Archirafi 36, I-90123 Palermo, Italy
E-mail: aurelio.agliologallitto@unipa.it

**Abstract**

In the paper we will describe three instruments of particular historical and didactic interest belonging to the Historical Collection of Physics Instruments of Palermo University: a convex mirror most likely dating back to the early nineteenth century and a pair of burning mirrors most likely dating back to the mid-nineteenth century. After an introduction on the historical development of mirrors and on their principle of operation, we will describe the characteristics of the instruments of the collection and the interventions carried out.

**Key words**: instruments of optics, burning mirrors, history of optics, museum of physics, scientific and technological heritage

**Riassunto**

Nell'articolo presenteremo tre strumenti scientifici di particolare interesse storico-didattico appartenenti alla Collezione Storica degli Strumenti di Fisica dell'Università di Palermo: uno specchio convesso risalente molto probabilmente all'inizio del XIX secolo e una coppia di specchi ustori risalenti molto probabilmente alla metà del XIX secolo. Dopo una introduzione sullo sviluppo storico degli specchi e il loro principio di funzionamento, descriveremo le caratteristiche fisiche degli strumenti della collezione e gli interventi effettuati.

**Parole chiave**: strumenti di ottica, specchi ustori, storia dell'ottica, museo di fisica, patrimonio scientifico e tecnologico

## Introduzione

L'Università degli Studi di Palermo custodisce un vasto patrimonio storico, artistico e scientifico di grande valore culturale. Con l'istituzione del Sistema Museale di Ateneo, nel 2011, di cui fanno parte sei musei tematici e quattordici collezioni dipartimentali (v. sito web 1), l'Università ha avviato un proficuo processo di tutela, conservazione e valorizzazione del proprio patrimonio culturale. La Collezione Storica degli Strumenti di Fisica (v. sito web 2) fa parte del Sistema Museale; essa è esposta, in ampie vetrine in legno dei primi del Novecento, al primo piano dell'edificio storico di via Archirafi 36 del Dipartimento di Fisica e Chimica - Emilio Segrè (Corrao, 2012; Spagnolo, 2019). La fig. 1 mostra una veduta dell'edificio di via Archirafi 36, sede dell'ex Istituto di Fisica e dell'ex Istituto di Mineralogia, subito dopo la sua costruzione intorno al 1930.

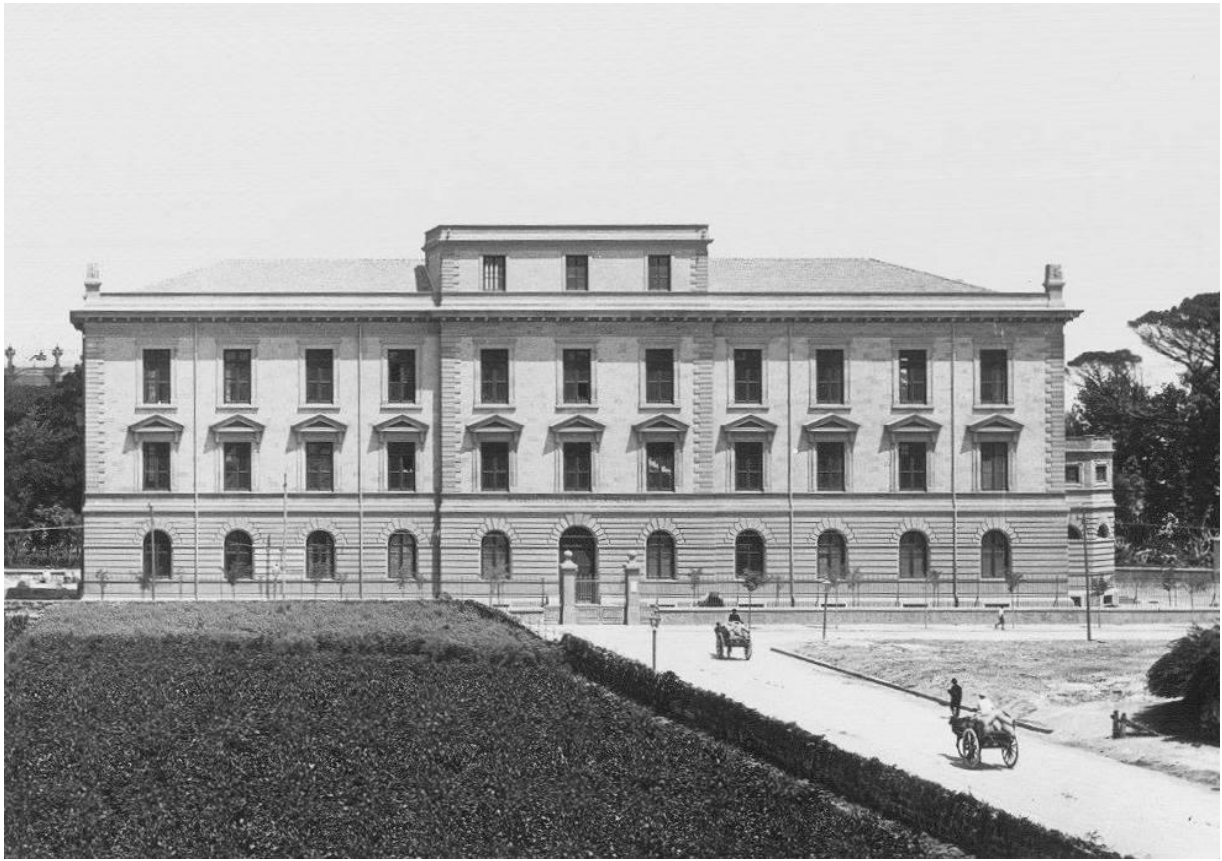

**Fig. 1.** Veduta dell'edificio storico di via Archirafi 36, sede dell'ex Istituto di Fisica e dell'ex Istituto di Mineralogia, subito dopo la sua costruzione intorno al 1930.

Fanno parte della collezione più di cinquecento strumenti e apparati scientifici di interesse storico-didattico (Sear, 2017; Agliolo Gallitto et al., 2018). I più antichi strumenti risalgono all'inizio del XIX secolo (Nastasi, 1998a e b), quando l'abate Domenico Scinà (1764 - 1837) divenne titolare della cattedra di Fisica Sperimentale, nel 1811. Un ritratto di

Domenico Scinà è mostrato nella fig. 2 (Scinà, 1840). In questo periodo, Scinà fece acquistare all'Università di Palermo svariati strumenti da usare a supporto della didattica (Nastasi, 1998a e b), la maggior parte di essi realizzati in officine locali. A questi primi strumenti didattici, nel corso del XIX e del XX secolo, si aggiunsero strumenti di ricerca. Sebbene, nel corso degli anni, per diversi motivi, molti di questi strumenti siano andati perduti, oggi la collezione comprende strumenti di meccanica, acustica, calorimetria, ottica, elettromagnetismo, spettroscopia e fisica moderna, a testimonianza degli interessi prevalenti della ricerca scientifica condotta a Palermo (Nastasi, 1998a e b; Russo, 1998; Agliolo Gallitto et al., 2017a e 2018).

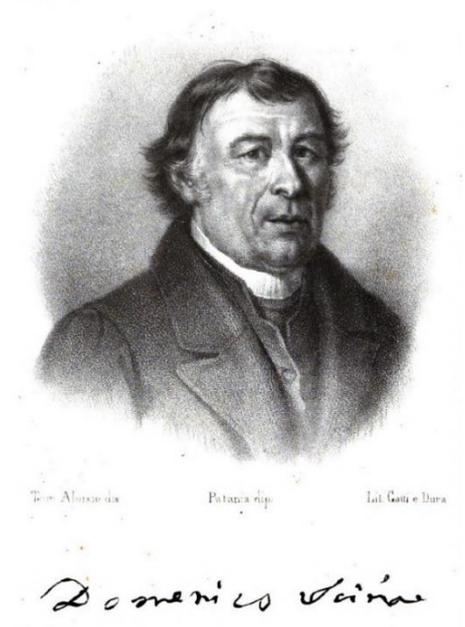

**Fig. 2.** Ritratto di Domenico Scinà (da Scinà, 1840).

Grazie al lavoro di recupero e conservazione condotto nel corso degli anni, prima dagli istituti universitari e oggi dai dipartimenti, le collezioni universitarie continuano ad arricchirsi. In questo contesto si inserisce il lavoro svolto, riguardante tre strumenti di ottica di particolare interesse storico-didattico: uno specchio convesso risalente molto probabilmente all'inizio del XIX secolo e una coppia di specchi ustori risalenti molto probabilmente alla metà del XIX secolo. Nell'articolo, dopo una introduzione sullo sviluppo storico degli specchi e sul loro principio di funzionamento, descriveremo le caratteristiche fisiche degli specchi della collezione e degli interventi effettuati.

## Sviluppo storico degli specchi e loro principio fisico

I primi specchi costruiti dall'uomo, costituiti da oggetti di ossidiana (vetro vulcanico) con superfici riflettenti, sono stati ritrovati in Anatolia (Turchia) in siti archeologici risalenti al 6000 a.C. (Enoch, 2006). A partire dal 4000 a.C., iniziarono a essere fabbricati specchi formati da dischetti di metallo lucidato, i quali ebbero grande diffusione nel periodo greco-romano e per tutto il medioevo, grazie allo sviluppo delle tecniche di lavorazione dei metalli (bronzo, argento e oro). Nel rinascimento ebbero grande diffusione gli specchi realizzati con lastre di vetro ricoperte di amalgama di stagno e mercurio, prodotte nelle botteghe veneziane. Nel XIX secolo, l'arte degli specchi veneziani iniziò la sua decadenza dopo l'introduzione del processo di argentatura dei vetri con soluzione di nitrato di argento, ammoniaca e acido tartarico (Arizio, 2011).

Dal punto di vista fisico, gli specchi hanno la proprietà di riflettere la luce. Essi sono caratterizzati da una superficie piana o da una superficie curva, principalmente sferica o parabolica. In particolare, gli specchi curvi possono essere concavi o convessi a seconda che la riflessione avvenga rispettivamente sulla superficie riflettente concava o su quella convessa. La riflessione della luce su uno specchio curvo si può dedurre dalla legge della riflessione per gli specchi piani, considerando la superficie dello specchio curvo formata da molte superfici piane infinitamente piccole. Gli specchi curvi sferici sono generalmente delle calotte sferiche di piccola apertura, tali da poter essere considerati con buona approssimazione dei paraboloidi.

Nell'antichità, gli specchi concavi furono oggetto di attenzione per la possibilità di impiego come specchi ustori. L'immaginario collettivo si è alimentato dalla vicenda di Archimede di Siracusa (287 a.C. circa - 212 a.C.) che usando un gran numero di specchi riuscì a bruciare le navi romane (Mills & Clift, 1992; Zamparelli, 2005; Acerbi, 2009). Non è chiaro, per gli storici, se questi specchi ustori siano stati veramente costruiti e utilizzati in guerra. Tuttavia, esperimenti recenti hanno mostrato che sarebbe possibile incendiare una nave, costruita con tecniche e materiali dell'antica Roma, concentrando opportunamente i raggi solari.

Una prima spiegazione del principio degli specchi ustori si trova nella Catottrica di Euclide (matematico greco del III secolo a.C.), il quale scrisse due libri di ottica: Ottica o Perspectiva (teoria della visione) e Catottrica (teoria delle immagini speculari). L'ultimo teorema della Catottrica è dedicato appunto al potere incendiario degli specchi ustori, risultando però inadeguato in quanto non indica nessuna costruzione del punto focale (Smith, 2001; Acerbi, 2009). Il matematico greco Diocle (III-II secolo a.C.) diede una dimostrazione rigorosa del loro funzionamento, dal punto di vista matematico. Egli fornisce la prima analisi conosciuta della proprietà focale degli specchi concavi parabolici, dimostrando che la parabola fa convergere tutti i raggi che la colpisco parallelamente al proprio asse in un unico punto, il fuoco. La dimostrazione di tipo geometrico si basa sul fatto che le tangenti nei punti di riflessione simulano uno specchio piano (Smith, 2001; Acerbi, 2009). Per un'approfondita trattazione storica dell'ottica e dello sviluppo degli strumenti ottici, si veda Calleri (2003) e Darrigol (2012). Nel corso dei secoli sono stati fatti diversi tentativi di ricostruzione dell'evento storico (Cavalieri, 1632). Tuttavia, poiché la distanza focale di uno specchio ustorio sferico si trova a metà del suo raggio, è ovviamente impossibile bruciare una sostanza situata a una grande distanza, poiché lo specchio dovrebbe essere di smisurata grandezza. Il padre gesuita tedesco Athanasius Kircher (1602 - 1680) pensò per primo di sostituire a uno specchio concavo più specchi piani inclinati e con cinque di essi esercitò un calore insopportabile alla distanza di 100 piedi (Scinà, 1833b). Nel XVIII secolo veniva usato il piede francese, che vale circa 0.325 m, e il piede inglese, che vale circa 0.305 m (Scinà, 1833a). Nel XVIII secolo, il matematico francese Georges-Louis Leclerc (1707 - 1788), conte di Buffon, con 400 specchi piani di mezzo piede quadrato, costituenti approssimativamente un paraboloide, riuscì a concentrare la luce del sole e far bruciare una tavola di legno ricoperta di catrame a una distanza di 200 piedi, fondere lo stagno a 150 piedi e il piombo a 141 piedi (Scinà, 1833b).

Gli specchi ustori venivano usati anche per mostrare che il "calorico raggiante" (radiazione infrarossa) si propaga e si riflette con le stesse leggi della luce (Agliolo Gallitto et al., 2016). Le leggi sulla propagazione del calorico raggiante erano state dedotte già nella seconda metà del XVIII secolo dai fisici svizzeri Marc Pictet (1752 - 1825) e Horace Bénédict de Saussure (1740 - 1799), e dal chimico svedese Carl Wilhelm Scheele (1742 - 1786). Pictet, in particolare, immaginò l'esperienza dei due specchi concavi messi a 24 piedi di distanza l'uno dall'altro, in cui nel fuoco dell'uno si poneva un carbone incandescente che infiammava un corpo combustibile posizionato nel fuoco dell'altro (Scinà, 1833b; Ganot, 1860). I raggi della sorgente vengono riflessi parallelamente all'asse principale e incidendo sul secondo specchio si focalizzano sul materiale combustibile procurandone l'accensione, come illustrato nella fig. 3 (v. sito web 3). In seguito, essi vennero impiegati anche per lo studio della propagazione del suono (Blaserna, 1875). Collocando una sorgente sonora nel fuoco di uno specchio, le onde sonore incidono sullo specchio e si riflettono verso il secondo specchio e da questo vengono concentrate nel proprio fuoco. Un orecchio collocato in questo punto sentirà distintamente il suono prodotto dalla sorgente posta nel fuoco del primo specchio.

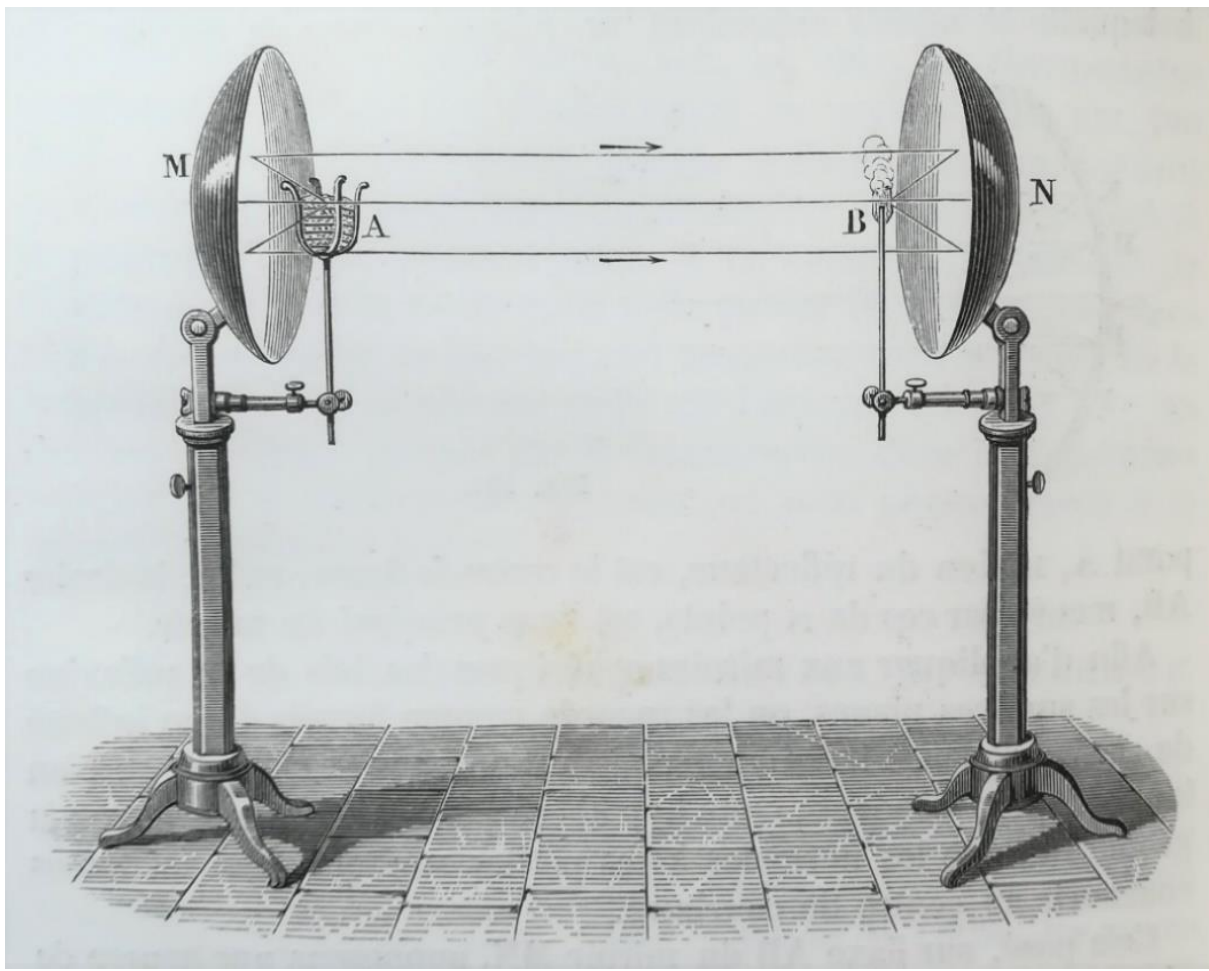

**Fig. 3.** Schema di funzionamento degli specchi ustori (da Ganot, 1860).

## Gli specchi della collezione

Lo specchio convesso della collezione era smontato e le vari parti erano conservate in differenti armadi. Dopo un'attenta ricognizione, è stato possibile riconoscere le tre parti costituenti lo strumento. In particolare, oltre allo specchio formato da una lastra di rame argentato, di diametro di 25 cm, inserita in una cornice di ottone, è stato ritrovato il piedistallo di ottone a tre piedi e l'arco semicircolare, sul quale è installato un meccanismo a molla con ruota dentata.

Sulle parti dello strumento non sono state rinvenute tracce della firma della casa costruttrice. Tuttavia, dalla consultazione di documenti di archivio, si riscontra che nell'inventario Scinà, antecedente al 1832, è riportato: "*Uno specchio concavo convesso di metallo con piede di rame*". Inoltre, dal confronto con altri strumenti della collezione, per affinità stilistica si può ipotizzare che lo specchio convesso sia stato realizzato molto probabilmente da "meccanici" locali all'inizio del XIX secolo, nel periodo in cui Domenico Scinà era titolare della cattedra di Fisica Sperimentale. D'altro canto, Scinà dedica un'ampia parte alla spiegazione del principio fisico dello specchio convesso nel suo trattato di fisica (Scinà, 1833b).

Lo strumento è stato quindi ricomposto nella sua forma originaria e è mostrato nella fig. 4. Lo specchio convesso agganciato all'arco semicircolare può essere posizionato con una inclinazione fissata e, a sua volta, posizionato sulla sommità del piedistallo di ottone tramite un innesto conico, può ruotare liberamente attorno all'asse verticale. Nella fig. 5, è mostrato in dettaglio il meccanismo a molla con ruota dentata per la regolazione dell'inclinazione dello specchio e il bloccaggio dello stesso in posizioni fisse e riproducibili. Considerato il buono stato di conservazione dello strumento, è stata effettuata una semplice pulitura delle superfici.

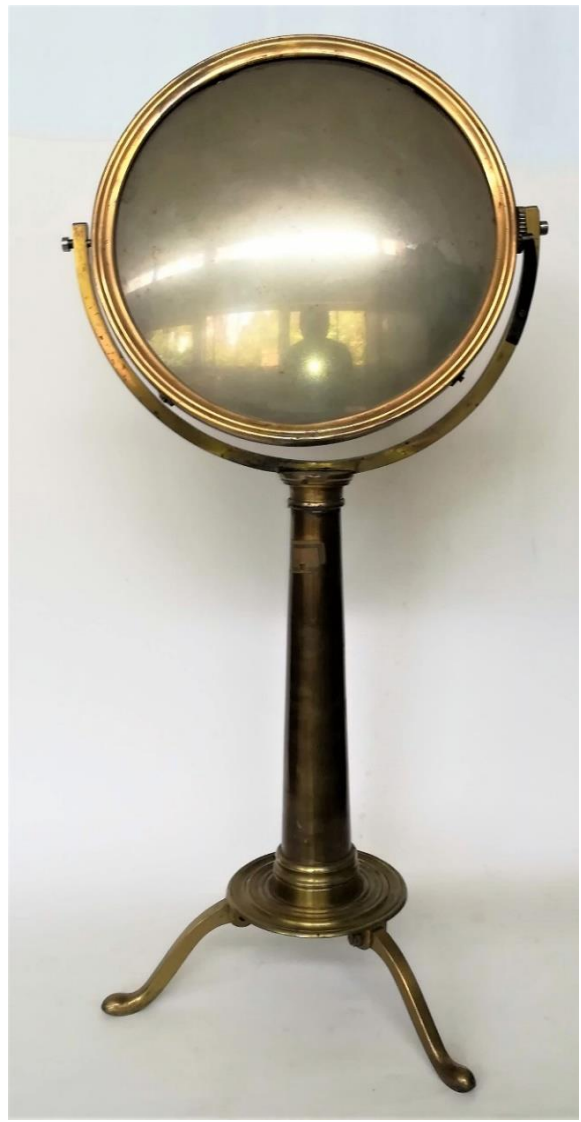

**Fig. 4.** Specchio convesso della collezione dell'Università di Palermo, ricomposto nella forma originaria.

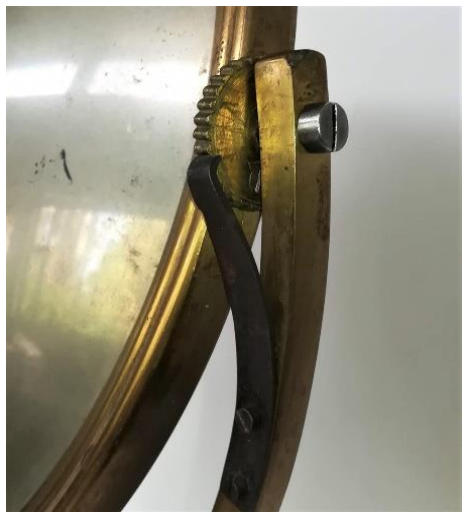

**Fig. 5.** Meccanismo a molla con ruota dentata per la regolazione dell'inclinazione dello specchio convesso e il bloccaggio dello stesso in posizione fisse e riproducibili.

Gli specchi concavi della collezione sono costituiti da un paraboloide di rame argentato di diametro di 43.5 cm, sostenuto da un piedistallo di ottone. Essi erano conservati incompleti: in uno specchio era mancante il piedistallo e nell'altro la base del piedistallo era priva della zavorra. Purtroppo, nonostante l'attenta ricognizione del materiale museale conservato, non sono state rinvenute tutte le parti mancanti ma solamente il piedistallo di ottone laccato privo della base e privo anche del raccordo a vite necessario per fissare il piedistallo allo specchio.

Sugli strumenti non sono state rinvenute tracce della firma della casa costruttrice, ma molto interessante è la consultazione dell'inventario Lo Cicero. Giuseppe Lo Cicero (1798 - 1887) ha prestato servizio presso l'Università di Palermo dal 1851 come dimostratore e dal 1857 come professore interino (provvisorio) di Fisica Sperimentale e Direttore del Gabinetto di Fisica fino al 1862 (Riccò, 1891). Dalla consultazione di documenti di archivio, nell'inventario Lo Cicero del 1857-1859 è stata individuata la seguente descrizione: "*Due specchi concavi di ottone pel calorico raggiante*". Tuttavia, è molto probabile che questi strumenti siano stati acquistati qualche anno prima della compilazione dell'inventario, in quanto Nastasi (1998a e b) riporta che nel 1847 fu ordinata "*l'ormai famosa termopila di Melloni*", in seguito alle lamentele circa l'esiguità della dotazione da parte degli studenti per potersi mettere in qualche modo al corrente dello stato attuale della scienza. Infatti, nello stesso periodo è stato acquistato, dalle officine Ruhmkorff di Parigi, un banco ottico del Melloni completo di accessori (Inv. N. 184 del 1850) che potrebbe essere identificato nella sopra citata "*termopila di Melloni*" e molto probabilmente nello stesso periodo furono acquistati gli specchi ustori per lo studio del "*calorico raggiante*". Inoltre, dal confronto con altri strumenti della collezione, si nota una marcata affinità stilistica con altri strumenti dello stesso periodo a firma Soleil/Duboscq. Nel XIX secolo, Soleil, Duboscq e Pellin hanno rappresentato una dinastia di costruttori francesi di strumenti scientifici (Brenni, 1996). Tuttavia, ulteriore indagine sui documenti storici di archivio è necessaria per poter determinare meglio la loro provenienza.

Sebbene uno specchio fosse privo della base, le varie parti degli strumenti nel complesso versavano in buono stato di conservazione. Pertanto, al fine di ricostituire l'integrità formale dell'apparato e potere esporre i due specchi nel loro insieme, si è deciso di integrare le parti mancanti. Questa decisione è stata presa anche sulla base del fatto che la ricostruzione non avrebbe apportato nessuna alterazione delle parti esistenti. Sono stati quindi costruiti il supporto e la base del piedistallo di ottone, mostrati nella fig. 6. Successivamente, la base di ottone è stata verniciata utilizzando una lacca protettiva accordata per colore e intensità a quella originale. Il supporto e le viti di ottone sono stati bruniti a freddo per via chimica. Si è proceduto quindi alla rimozione del deposito superficiale incoerente, con l'uso di pennellesse a setole morbide, e successivamente alla rimozione meccanica, con spatola di legno, del deposito superficiale coerente (incrostazioni di intonaco/calcestruzzo), sul retro di uno dei due specchi, dopo ammorbidimento con acqua demineralizzata applicata localmente a tampone (fig. 7). Quindi, sono stati puliti gli elementi di ottone laccato e lucidate le superfici argentate.

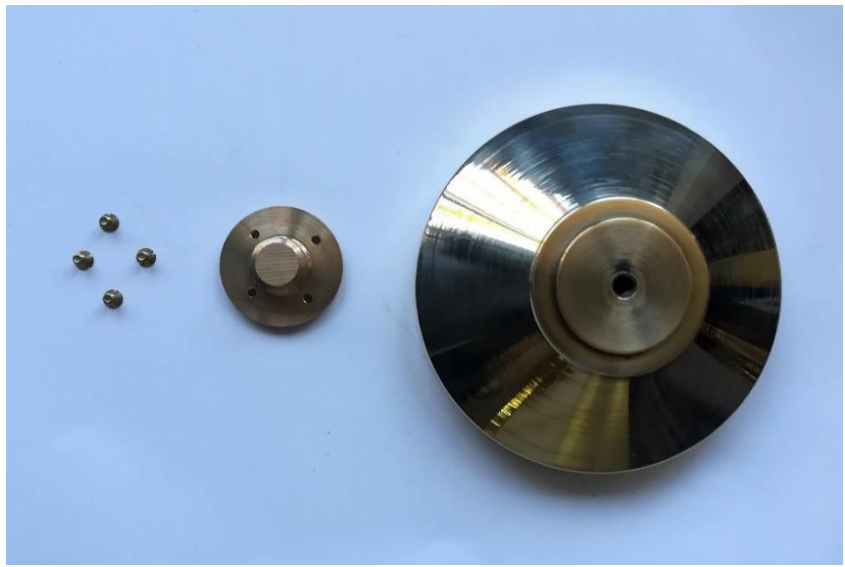

**Fig. 6.** Parti di ottone ricostruite.

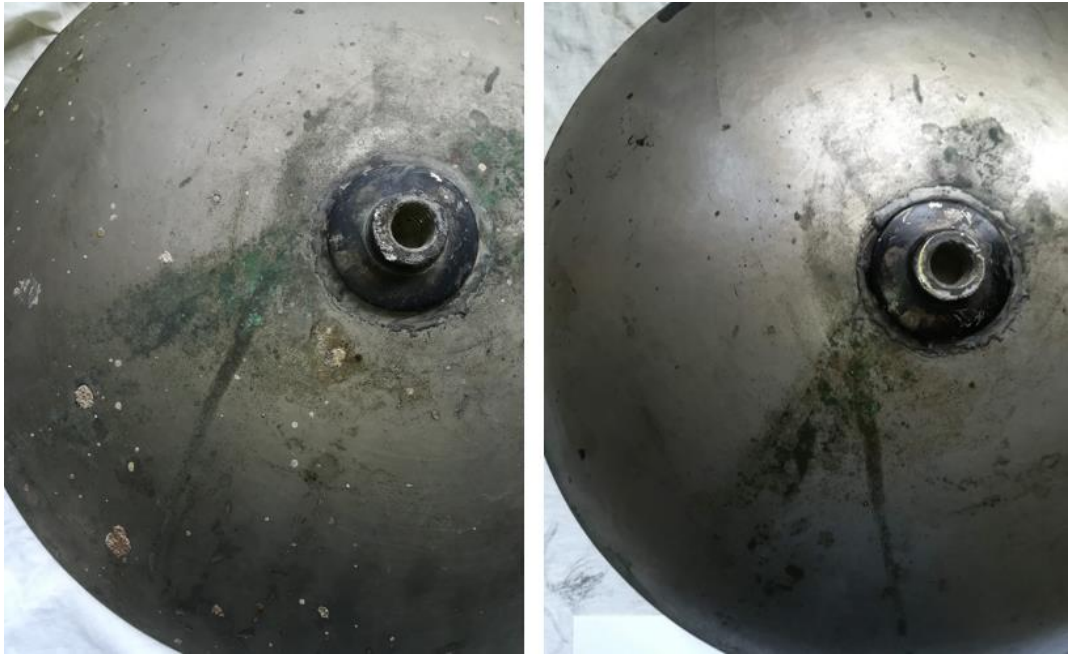

**Fig. 7.** Immagine della superfice posteriore dello specchio concavo prima (sinistra) e dopo (destra) la rimozione dei depositi superficiali.

La realizzazione delle parti mancanti è stata effettuata in modo da rendere ben riconoscibili le parti ricostruite a una attenta ispezione. Vale la pena sottolineare che l'intervento effettuato è pienamente reversibile e che le parti ricostruite potranno essere sostituite qualora si venisse in possesso degli originali. Inoltre, per rendere più stabili gli specchi nell'esposizione, sono stati realizzati due supporti di plexiglas che opportunamente posizionati ne impediscono il ribaltamento. La fig. 8 mostra la foto degli specchi ustori dopo gli interventi effettuati. Infine, entrambi gli apparati sono stati catalogati, compilando schede descrittive, con le voci essenziali per una successiva integrazione nella scheda PST (Patrimonio Scientifico e Tecnologico) dell'Istituto Centrale per il Catalogo e la Documentazione (Miniati, 2008), e collocati nelle vetrine espositive della collezione, garantendo in questo modo un ambiente stabile per la loro conservazione.

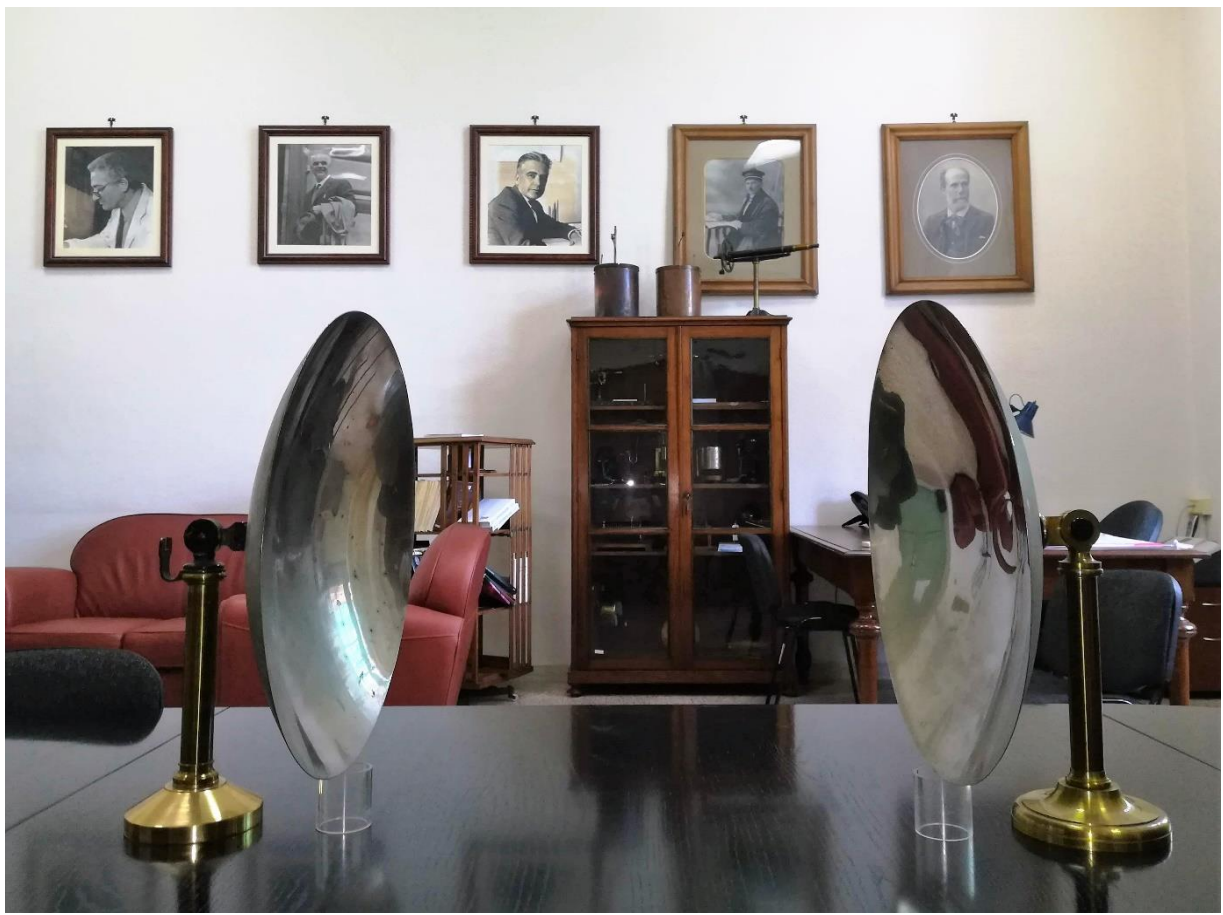

**Fig. 8.** Specchi ustori della collezione dell'Università di Palermo, dopo gli interventi di recupero.

## Discussione e conclusioni

Negli ultimi decenni, i musei scientifici hanno avuto un ruolo sempre più importante nello sviluppo culturale, non solo per gli studenti delle scuole (Pantano & Talas, 2010; Barbacci et al., 2012; Agliolo Gallitto et al., 2017b) ma anche per il pubblico in generale (Mujtaba, 2018). Sebbene i musei scientifici pubblici si evolvano verso una più ampia missione culturale consentendo ai visitatori di esplorare in modo interattivo i fenomeni scientifici, i musei scientifici universitari mantengono un loro ruolo originale, che è quello di preservare la storia e il patrimonio culturale legato allo sviluppo della disciplina specifica. Gli strumenti della Collezione Storica degli Strumenti di Fisica dell'Università di Palermo costituiscono pertanto un interessante mezzo per la ricostruzione dello sviluppo della fisica nei suoi aspetti storici e didattici. Inoltre, l'approccio allo studio delle leggi della fisica basato sullo studio degli strumenti storici può essere rilevante anche da un punto di vista didattico, in quanto gli strumenti storici e i relativi esperimenti scientifici possono essere usati per rafforzare una comprensione autentica delle leggi della fisica e guidare gli studenti verso una completa e coerente conoscenza scientifica (Metz & Stinner, 2007; Agliolo Gallitto et al., 2021).

Gli interventi sugli strumenti della collezione, considerato il loro pregio, sono stati effettuati seguendo il principio del minimo intervento, della riconoscibilità e reversibilità dell'intervento e della compatibilità dei materiali usati con quelli originali (Brandi, 1977; Giatti & Miniati, 1988; Marotti, 2004). Lo stato di conservazione degli strumenti testimoniava lo stato di abbandono sofferto soprattutto nel corso del secolo scorso: dalle condizioni generali delle superfici con l'annerimento sia delle superfici argentate, sia della vernice di protezione dell'ottone, alla perdita di parti meccaniche. Particolare attenzione è stata rivolta alla lacca protettiva dell'ottone cercando di preservarla, effettuando solo una accurata pulitura. Infatti, la lacca rappresenta una delle principali caratteristiche peculiari della manifattura degli strumenti scientifici (Lanterna & Giatti, 2014; Brenni et al., 2019). Gli interventi sono stati effettuati con l'obiettivo di preservare tutti gli elementi e, nello stesso tempo, recuperare l'aspetto originario per consentire la lettura di ogni oggetto nel suo insieme, tenendo conto della sua storia, anche conservativa, e dello stato di conservazione attuale, cercando di trovare un punto di equilibrio fra il recupero delle funzioni dello strumento e il rispetto del suo essere documento storico. Tali interventi hanno consentito di ottenere una idonea musealizzazione, facilitare la comprensione del principio di funzionamento degli strumenti da parte del pubblico meno esperto e contribuire in questo modo alla valorizzazione degli strumenti stessi e della collezione a cui essi appartengono (Giatti & Miniati, 1988; Marotti, 2004; Agliolo Gallitto et al., 2016; Brenni et al., 2019).

In conclusione, nell'articolo sono stati descritti tre strumenti scientifici di particolare interesse storico-didattico appartenenti alla Collezione Storica degli Strumenti di Fisica dell'Università di Palermo: uno specchio convesso risalente molto probabilmente all'inizio del XIX secolo e una coppia di specchi ustori risalenti molto probabilmente alla metà del XIX secolo. Inoltre, sono stati discussi gli aspetti storico-didattici legati allo studio delle leggi fisiche che sono alla base del principio di funzionamento di questi strumenti.

## Ringraziamenti

# Bibliografia

ACERBI F., 2009. *I geometri greci e gli specchi ustori*. In: *Matematica, cultura e società 2007–2008*, Ed. Gabbani, Pisa, pagg. 187–230 (www.academia.edu/8016440/I_geometri_greci_e_gli_specchi_ustori).

AGLIOLO GALLITTO A., LICATA S., MIRABELLO F., TAORMINA F., 2016. Recupero di un raro banco ottico del Melloni costruito nella Palermo della "belle époque", *Museologia Scientifica*, **10**: 117–121.

AGLIOLO GALLITTO A., CHINNICI I., BARTOLONE F., 2017a. *Collezione Storica degli Strumenti di Fisica: Catalogo degli strumenti di Acustica.* Università degli Studi di Palermo, Palermo (sites.google.com/site/aurelioagliologallitto/collezione-storica/ebook_acustica).

AGLIOLO GALLITTO A., PACE V., ZINGALES R., 2017b. Multidisciplinary learning at the university scientific museums: the Bunsen burner, *Museologia Scientifica*, **11**: 103–107.

AGLIOLO GALLITTO A., CHINNICI I., BARTOLONE F., 2018. Gli Strumenti di Acustica della Collezione Storica degli Strumenti di Fisica dell'Università di Palermo. *Museologia Scientifica*, **12**: 48–54.

AGLIOLO GALLITTO A., ZINGALES R., BATTAGLIA O. R., FAZIO C., 2021. An approach to the Venturi effect by historical instruments, *Physics Education*, **56**: 025007–9 (doi: 10.1088/1361-6552/abc8fa).

ARIZIO L., 2011. *Specchi dal XVII secolo ai giorni nostri: studio chimico fisico preliminare su vetri, strati riflettenti e loro degrado.* Tesi di Laurea Magistrale in Scienze Chimiche per la Conservazione ed il Restauro, Università Ca' Foscari A.A. 2011/2012, Venezia.

BARBACCI S., BRENNI P., GIATTI A., 2012. *Strumenti scientifici: object reading e didattica informale*. In: Peruzzi A., (a cura di). *Pianeta Galileo 2011*. Consiglio regionale della Toscana, Firenze, pagg. 183–198.

BLASERNA P., 1875. *La teoria del suono nei suoi rapporti colla musica*. Dumolard, Milano (books.google.it/books?id=9-ksAAAAYAAJ).

BRANDI C., 1977. *Teoria del restauro*, Einaudi, Torino.

BRENNI P., 1996. 19th Century French Scientific Instrument Makers. XIII: Soleil, Duboscq, and Their Successors. *Bulletin of the Scientific Instrument Society*, **51**: 7–16.

BRENNI P., GIATTI A., SERRA L., VALLE M., 2019. Valorizzazione di una collezione scientifica: museo e istituzioni collaborano, *Museologia Scientifica Memorie*, **19**: 25–29.

CALLERI M., 2003. *Lineamenti di storia dell'ottica: dalle lenti ustorie ai laser*. Università degli Studi di Torino, Torino (dattiloscritto).

CAVALIERI B., 1632. *Lo specchio ustorio*. Clemente Ferroni, Bologna (gallica.bnf.fr/ark:/12148/bpt6k51250b).

CORRAO R., 2012. *Architettura e Costruzione nella Palermo tra le due Guerre. Tre edifici pubblici emblematici.* Aracne, Roma, pp. 19–28.

DARRIGOL O., 2012. *A History of Optics: From Greek Antiquity to the Nineteenth Century.* Oxford University Press, Oxford.

ENOCH J. M., 2006, History of Mirrors Dating Back 8000 Years. *Optometry and Vision Science*, **83**: 775–781 (doi: 10.1097/01.opx.0000237925.65901.c0).

GANOT A., 1860. *Traité élémentaire de physique expérimentale et appliquée et de météorologie.* Chez L'Auteur – Éditeur, Paris.

LANTERNA G., GIATTI A., 2014. Caratterizzazione non invasiva delle vernici da ottone degli strumenti scientifici: Ricette storiche, realizzazione di provini verniciati, ricerca analitica e applicazioni "in situ" su strumenti storici. *OPD Restauro*, **26**: 165–180 (www.jstor.org/stable/24398171).

MAROTTI R., 2004. *Introduzione al restauro della strumentazione di interesse storico scientifico*. Il Prato, Padova.

METZ D., STINNER A., 2007. A Role for Historical Experiments: Capturing the Spirit of the Itinerant Lecturers of the 18th Century. *Science & Education*, **16**: 613–624.

MILLS A. A., CLIFT R., 1992. Reflections of the 'Burning mirrors of Archimedes'. With a consideration of the geometry and intensity of sunlight reflected from plane mirrors. *European Journal of Physics*, **13**: 268–279 (doi : 10.1088/0143-0807/13/6/004).

GIATTI A., MINIATI M. (a cura di), 1988. *Il restauro degli strumenti scientifici*. Alinea Editrice, Firenze.

MINIATI M., 2008. Catalogazione di strumenti scientifici: dalla scheda STS alla scheda PST, *Museologia Scientifica Memorie*, **2**: 18–20.

MUJTABA T., LAWRENCE M., OLIVER M., REISSA M. J., 2018. Learning and engagement through natural history museums. *Studies in Science Education*, **54**: 41–67.

NASTASI P., 1998a, Domenico Scinà e la fisica palermitana fra Settecento e Ottocento. *Studi Settecenteschi*, **18**: 377–405.

NASTASI P., 1998b. Da Domenico Scinà a Michele La Rosa, Le scienze chimiche, fisiche e matematiche nell'ateneo di Palermo. In: Nastasi P. (a cura di) *Quaderni del Seminario di Storia della Scienza* N. 7, Cap. 1, pagg. 121–165, Università di Palermo, Palermo.

PANTANO O. and TALAS S., 2010. Physics thematic paths: laboratorial activities and historical scientific instruments. *Physics Education*, **45**: 140–146 (doi: 10.1088/0031-9120/45/2/002).

RICCÒ A., 1891. Sulla vita e sulle opere del defunto socio Cav. Prof. Giuseppe Lo Cicero. In: *Atti della Reale Accademia di Scienze, Lettere e Belle Arti di Palermo*, Palermo (archive.org/details/attidellaaccadem0301acca/)

RUSSO A., 1998. L'affermazione della fisica palermitana nel panorama scientifico nazionale, 1935-1970. In: Nastasi P. (a cura di) *Quaderni del Seminario di Storia della Scienza* N. 7, Cap. 2, pagg. 167–193, Università di Palermo, Palermo.

SCINÀ D., 1833a. *Elementi di fisica generale*. Tomo I, Società tipografica de' classici italiani, Milano (archive.org/details/bub_gb_TJSYZPcBJKYC).

SCINÀ D., 1833b. *Elementi di fisica particolare*. Tomo I, Società tipografica de' classici italiani, Milano (archive.org/details/bub_gb_8Y5JJarUjE8C).

SCINÀ D., 1840. *Storia letteraria di Sicilia ne' tempi greci*. Tipografia Trani, Napoli (archive.org/details/bub_gb_KvIrAAAAYAAJ_2).

SEAR T., 2017. The Historical Collection of Physics Instruments of Palermo University, *Bulletin of the Scientific Instrument Society*, **132**: 32–33.

SMITH M., 2001. *Scienza greco-romana. Ottica e teoria della luce*. Storia della Scienza, Enciclopedia Treccani.

SPAGNOLO B., 2019. 1937, Palermo: the discovery of technetium. *SIF PrimaPagina* N. **62**, febbraio (www.primapagina.sif.it/article/911/1937-palermo-the-discovery-of-technetium).

ZAMPARELLI C., 2005. Storia, Scienza e Leggenda degli specchi ustori di Archimede. *Didattica delle scienze*, **XXIII**, 193, pagg. 52-56.

**Siti web**

1. *Sistema Museale di Ateneo dell'Università degli Studi di Palermo*
   musei.unipa.it

2. *Collezione Storica degli Strumenti di Fisica dell'Università degli Studi di Palermo*
   sites.google.com/site/aurelioagliologallitto/collezione-storica

3. Canale youtube della Fondazione Scienza e Tecnica di Firenze, *Pair of parabolic mirrors for heat reflection* (a cura di Giatti A. e Brenni P.)
   youtu.be/aJMSBf6frtw